\documentclass[aps,prl,twocolumn,groupedaddress]{revtex4-1}

\usepackage{graphicx}
\usepackage{epsfig}
\usepackage{fancyhdr}
\usepackage{amsmath,amsfonts,amssymb,amsthm}

\def\cG{{\mathcal{G}}}

\def\cR{{\mathcal{R}}}

\def\be{\begin{equation}}
\def\ee{\end{equation}}
\def\ben{\begin{equation*}}
\def\een{\end{equation*}}

\newtheorem{thm}{Theorem}
\newtheorem{cor}{Corollary}

\newtheorem{prop}{Proposition}

\begin{document}


\title{Transport and Vulnerability in River Deltas: A Graph-Theoretic Approach}


\author{Alejandro Tejedor}
\affiliation{St. Anthony Falls Laboratory and National Center for Earth-surface Dynamics, University of Minnesota, Minneapolis, MN, USA 55414}
\author{Anthony Longjas}
\affiliation{St. Anthony Falls Laboratory and National Center for Earth-surface Dynamics, University of Minnesota, Minneapolis, MN, USA 55414}
\author{Ilya Zaliapin}
\affiliation{Department of Mathematics and Statistics, University of Nevada, Reno, NV, USA 89557}

\author{Efi Foufoula-Georgiou}
\affiliation{Department of Civil Engineering, University of Minnesota, Minneapolis, MN, USA 55414}


\date{\today}

\begin{abstract}
Maintaining a sustainable socio-ecological state of a river delta requires 
delivery of material and energy fluxes to its body and coastal zone in a 
way that avoids malnourishment that would compromise system integrity.  
We present a quantitative framework for studying delta topology and transport based on representation of a deltaic system by a rooted directed acyclic graph.
Applying results from spectral graph theory allows systematic 
identification of the upstream and downstream subnetworks for a 
given vertex, computing steady flux propagation in 
the network, and finding partition of the flow at any channel among the 
downstream channels.
We use this framework to construct vulnerability maps that quantify 
the relative change of sediment and water delivery to the shoreline outlets 
in response to possible perturbations in hundreds of upstream links.  
This enables us to evaluate which links (\emph{hotspots}) and what management 
scenarios would most influence flux delivery 
to the outlets.  
The results can be used to examine local or spatially distributed delta 
interventions and develop a system approach to delta management.   
\end{abstract}


\maketitle

Deltas are landforms with channels that deliver water, sediment and nutrient 
fluxes from rivers to oceans or inland water bodies via multiple pathways.  
These systems evolve naturally by maintaining the balance between subsidence 
due to compaction and new land formation due to sediment deposition from the river upstream~\citep{Paola11}.  This dynamic interaction results in a low-relief terrain with slopes as low as 1.0 x 10$^{-5}$~\citep{Syvitski05}.  

Deltas are highly productive regions supporting extensive agriculture, diverse ecosystems, and containing natural resources such as hydrocarbon deposits.  
More than half a billion people reside in deltas with over 300 million 
living in the Ganges-Bramahputra-Meghna, Yangtze and Nile alone~\citep{Syvitski09,Foufoula-Georgiou13}.  Unfortunately, many deltas are vulnerable to both natural and anthropogenic drivers, and are predicted to be in danger of collapsing within the 21st century~\citep{Syvitski09}.  Any alteration in the delta network can provoke physical (channel morphology), biological (ecosystems) and socio-economic changes.  For instance, rapid sea-level rise exacerbates land loss inducing saltwater intrusion upstream and loss of ecosystem habitat. 
On the other hand, demands on water and energy from upstream are satisfied 
by dams and divergence structures, while multiple dykes, embankments and 
sluice gates are constructed downstream mainly for irrigation purposes 
and to control floods and salinity intrusion.  These~\emph{human-engineered} structures result in unintended consequences that prevent the growth of deltas and disrupt their natural ecosystem dynamics.  The rising sea-level, in concert with reduction in aggradation and accelerated compaction, is putting many deltas in peril~\citep{Syvitski09,BlumRoberts09}.   

Recent works have focused mainly on modeling delta growth and 
evolution~\citep{JerolmackSwenson07, Seybold07,F08,EdmondsSlingerland09, Wolinsky10,Shaw13} 
and developing quantitative metrics to describe delta morphology~\citep{Edmonds11,Passalacqua13}.  At the same time, formal methodologies for studying deltas' topology and dynamic 
processes operating on them are still lacking.  
Our study contributes to this direction.  
Specifically, we conceptualize a delta as a rooted acyclic directed graph
and use its weighted adjacency matrix to 
(i) identify the steady flow along the delta channels, 
(ii) detect upstream (contributing) and downstream (nourishing) 
subnetworks for a given channel,
(iii) find the partition of flow from a given channel to
any collection of downstream nodes at steady state, and
(iv) examine how perturbations at upstream parts of the system propagate downstream.
The developed framework is illustrated by examining the vulnerability of the downstream (shoreline) 
outlets to flux reduction in upstream channels and building $\emph{vulnerability maps}$ 
that can facilitate assessment of delta development scenarios.  

\begin{figure}
\centering\includegraphics[width=.45\textwidth]{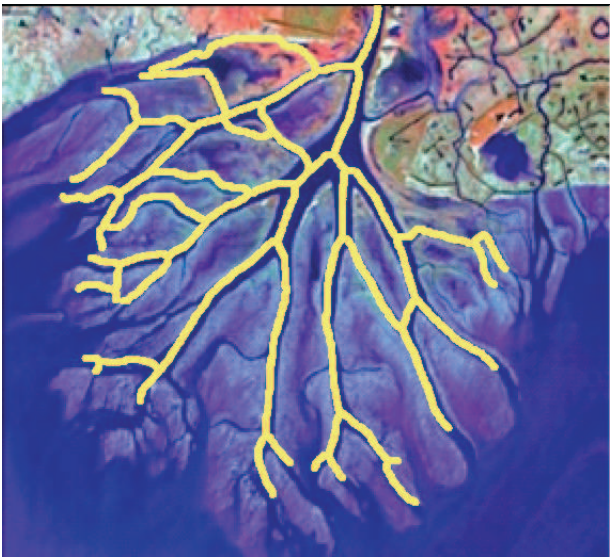}
\caption{Wax Lake delta. 
The skeleton network (yellow lines) is superimposed on the aerial view of the delta in 2005 by the National Center of Earth-surface Dynamics (NCED).  
Only links in the network that connect the delta apex to the shoreline outlets are considered in the connectivity analysis.}
\label{fig:fig1}
\end{figure}
 


Recall that a graph $\cG=(V,E)$ is a collection of vertices $V=\{v_i\}$, $i=1,\dots, N$ 
and edges $E=\{(uv)\}$, $u,v\in V$, where the notation $(uv)$ signifies that the edge
connects the vertices $u$ and $v$. 
A graph is called {\it directed}, or {\it digraph}, if its edges have directions, that 
is, edge pairs $(uv)$ are ordered.
A digraph is called {\it acyclic} if there are no directed paths from a vertex to itself.  
A digraph is called {\it rooted} if there is a vertex $r$ such that there
exists a directed path from $r$ to any other vertex in the graph.
A digraph can be uniquely specified by its (asymmetric) {\it adjacency matrix} $A$ such that $A(v,u)=1$ if there is an edge $(uv)$ and $A(v,u)=0$ otherwise.
Sometimes, edges are given weights $w_{uv}$. 
In this case the graph is specified by its {\it weighted} adjacency matrix $W$ with
non-zero elements $W(v,u)=w_{uv}$.
In a {\it marked} graph $\cG=\{V,E,F\}$ each vertex has a quantitative 
characteristic $F_i$. 
An acyclic digraph imposes well defined parent-child relationships among 
the vertices.
Specifically, each edge $(uv)$ connects a {\it parent} $u$ to a {\it child} (offspring) $v$.
In general, each vertex may have multiple offspring and parents.
The vertices with no offspring are called {\it leaves}.

We consider all the links in the delta network that connect the apex to the 
shoreline outlets~\citep{Edmonds11}, and
assume unique downstream direction of fluxes along the delta channels. 
The topological arrangement of the delta channels can be represented by 
a marked rooted acyclic directed graph $\cG$.
The delta apex corresponds to the root of $\cG$; 
the shoreline outlets -- to the leaves;
the physical points where channels intersect (combine or split) or terminate
-- to vertices; and the channel segments between intersections/splits/outlets -- to edges.
The direction of flux through the delta (from root to outlets) is represented 
by the edge directions.
The flux intensity at node $i$ is given by time-dependent mark $F_i(t)$.
The distribution of the flux at a parent vertex $u$ among the offspring vertices 
$(x,\dots,z)$ is given by vector $(w_{ux},\dots,w_{uz})$ such that 
$w_{ux}+\dots+w_{uz}=1$. 
The weights $w_{uv}$ form the weighted adjacency matrix $W$. 

The problem of finding the steady flow on a graph is 
well-known in transport network studies. 
At steady state, the flow through vertex $i$ equals the
total in-flow from its parents:
$F_i = \sum_{j}w_{ji}F_j.$
This equation applies to all the vertices except the root (which does not have
the source of flux) and leaves (which do not have offspring) -- at these vertices
the flux cannot reach a nontrivial steady state.
To avoid this problem we consider a {\it cycled} version of the network, where
the outlets (leaves) directly drain their entire flux to the apex (root). 
The weighted adjacency matrix for this new network is denoted by
$\tilde W$.
Hence, we seek a solution $F=(F_1,\dots,F_N)^{\rm T}$ 
(where $x^{\rm T}$ is the transpose of $x$) of the system
\be
F_i = \sum_{j}\tilde w_{ji}F_j, \quad i=1,\dots,N.
\ee
This can be written in matrix notation as $F = \tilde W\,F$,
or $({\bf I}_N-\tilde W)F={\bf 0}_N$, where ${\bf I}_N$ is the $N\times N$ 
identity matrix, and ${\bf 0}_N$ is an $N\times 1$ 
vector of zeros.
In other words, we need to identify the null space of the matrix
$L_{\rm SF}={\bf I}-\tilde W$.
We also notice that ${\bf I}_N$ is the out-degree matrix for $\tilde W$, so
$L_{\rm SF}$ is a graph Laplacian for $\tilde W$.
The steady flow at an edge $(uv)$ is given by $F_{(uv)}=F_u\,w_{uv}$.  
It is clear from the flux interpretation that there exists a unique solution
to this problem.

\begin{prop}[Steady flow]
\label{SF}
The steady flow $F$ through a rooted acyclic digraph $\cG$
with weighted cycled adjacency matrix $\tilde W$ is given (up to
a scalar factor) by the eigenvector that spans the null space of the
graph Laplacian:
\ben
{\rm null}\left(L_{\rm SF}\right)={\rm null}\left({\bf I}-\tilde W\right)
=\left\{x:({\bf I}-\tilde W)x={\bf 0}_N\right\}.
\een  
\end{prop}

Next, we use the results of \cite{AC05,CV06} to identify 
the subnetwork that participates in draining fluxes from 
the apex to a given vertex $u$ (contributing network)
and the subnetwork that drains fluxes from $u$ to the outlets
(nourishing network) \citep{Edmonds11}. 
Recall that a {\it reachable} set $\cR(i)$ for vertex $i$ 
in a digraph $\cG$ is the collection of all vertices $j$
such that there exists a directed path from $i$ to $j$.
The set $\cR$ is called a {\it reach} if it is a maximal reachable set,
that is $\cR=\cR(i)$ for some $i$ and there is no such $j$ that 
$\cR(i)\subset \cR(j)$.
The {\it exclusive part} of a reach $\cR_i$ is defined as 
$H_i=\cR_i\setminus \cup_{j\ne i}\cR_j$.
The {\it common part} of $\cR_i$ is defined as $C_i=\cR_i\setminus H_i$.
Caughman and Veerman \cite{CV06} prove the following theorem.

\begin{thm}[Reaches of a digraph]
Let $A$ be the $N\times N$ adjacency matrix for $\cG$ and $D$ 
be the in-degree matrix for $A$, that is the diagonal $N\times N$ 
matrix with diagonal elements taken from $A{\bf 1}_N$. 
Then the nullspace of the Laplacian $L=D-A$ has a basis
$\gamma_i$ in $\mathbb{R}^N$ whose elements satisfy:
(i) $\gamma_i(v)=0$ for $v\notin\cR_i$;
(ii) $\gamma_i(v)=1$ for $v\in H_i$;
(iii) $\gamma_i(v)\in(0,1)$ for $v\in C_i$;
$\sum_i\gamma_i = {\bf 1}_N$.
\end{thm}

Suppose that a delta system is represented by a rooted acyclic digraph $\cG$. 
Consider now the same stream topology with {\it reversed} flux directions; 
the new network is specified by the acyclic directed graph $\cG^{\rm R}$ with 
adjacency matrix $A^{\rm T}$.
Each outlet $i=1,\dots,k$ of the initial delta generates a reach $\cR_i$  
within $\cG^{\rm R}$.
The exclusive part of $\cR_i$ consists of the vertices that 
in $\cG$ drain exclusively to the outlet $i$.
The common part of $\cR_i$ consists of the vertices that
in $\cG$ also drain to at least one other outlet.
It is easily seen that there are no other reaches in $\cG^{\rm R}$.
Each outlet $i$ of the initial delta belongs to the exclusive part
of the respective reach in $\cG^{\rm R}$.
The apex of the initial delta belongs to all the reaches of $\cG^{\rm R}$.
Let $B$ denote the in-degree matrix for $A^{\rm T}$, that is the diagonal
$N\times N$ matrix with diagonal elements taken from $A^{\rm T}{\bf 1}_N$.
Consider the Laplacian $Q=B-A^{\rm T}$of the digraph $\cG^{\rm R}$.
A basis for the nullspace of $Q$ is described by Theorem 1.
The subnetwork of $\cG$ that drains from the apex to the 
outlet $i$ corresponds to the unique eigenvector of the $Q$-nullspace
with $\gamma(i)=1$.
The nonzero elements of this eigenvector identify the vertices that
participate in the subnetwork.

Notably, if we consider a flow along the directed edges of $\cG$ with 
equal distribution of the parental flux among the offsprings, then
the elements of eigenvectors from the $Q$-nullspace allow the following interpretation. 
The element $\gamma_i(v)$ equals the proportion of the flux
at vertex $v$ that drains to outlet $i$ in the original network $\cG$.
This statement is trivial for the outlets of $\cG$, which always 
belong to the exclusive part of respective reaches. 
To prove the statement for the rest of vertices,
observe that $B\gamma_i=A^{\rm T}\gamma_i$, which means 
that the element $v$ of the eigenvector $\gamma_i$ 
equals the average of its parental elements in $\cG^{\rm R}$.
The results of \cite{CV06} apply as well to general stochastic adjacency 
matrices, which allows one to generalize the above discussion to the 
situation with unequal distribution of parental flux among the offsprings. 

\begin{cor}[Contributing network]
\label{reach}
Consider flow along the edges of a rooted acyclic graph $\cG$
specified by a weighted adjacency matrix $W$ whose elements $w_{uv}$ 
represent the proportion of flux at parental vertex $u$ that drains 
to offspring vertex $v$.
Let $Q=B-W^{\rm T}$ be the weighted graph Laplacian,
with $B$ being the in-degree matrix for $W^{\rm T}$.
Then the nullspace of $Q$ has a basis
$\gamma_i$ in $\mathbb{R}^N$ such that $\gamma_i(v)$ equals the
proportion of flux at vertex $v$ that drains to outlet $i$.
In particular, $\gamma_i(v)\ne 0$ if and only if outlet $i$
receives fluxes from vertex $v$. 
\end{cor}

Corollary~\ref{reach} is readily applied to
finding the subnetwork that participates in draining fluxes to
any chosen vertex, not necessarily an outlet.
For that, one needs to make the examined vertex an outlet,
by disconnecting it from its offspring.
The modified weighted adjacency matrix is then used to obtain
the result.

To identify the nourishing network for a node $u$,
we first make $u$ an apex by disconnecting it from the parents.
The identification is done using the following result.

\begin{cor}[Nourishing network]
\label{down}
Consider flow along the edges of an acyclic graph $\cG$ with
$k$ roots, specified by a weighted adjacency matrix $W$ 
whose elements $w_{uv}$ represent the proportion of flux at parental 
vertex $u$ that drains to offspring vertex $v$.
Let $L=D-W$ be the weighted graph Laplacian 
with $D$ being the in-degree matrix for $W$.
Then the nullspace of $L$ has a basis
$\gamma_i$, $i=1,\dots,k$ in $\mathbb{R}^N$ such 
that $\gamma_i(v)\ne 0$ if and only if vertex $v$ receives 
fluxes from root $i$.
\end{cor}

\begin{figure}[ht]
\includegraphics[width=.45\textwidth]{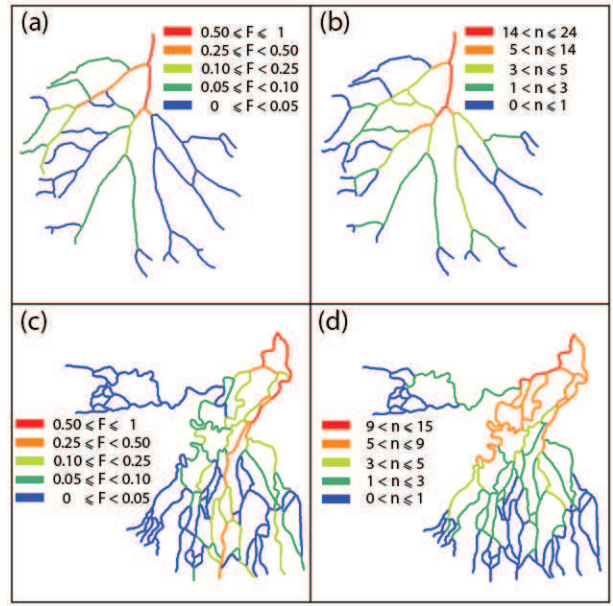}
\caption{Steady state flux (a,c) and number of outlets (b,d) that a given link 
contributes to.
(a,b) Wax Lake delta: The distribution of flux among the immediate downstream links
is proportional to the channel width.
(c,d) Niger delta: The flux is distributed equally among the immediate downstream links.
The flux at the apex is normalized to $F=1$.}
\label{fig:fig2}
\end{figure}



We illustrate this framework with the Wax Lake and Niger deltas.  
The Wax Lake delta is a relatively young river-dominated delta with a 
radial shoreline propagation (Fig.~\ref{fig:fig1})~\citep{Paola11}.  
It directly receives input from the Wax Lake outlet with an average 
discharge of 2,783 $m^{3}/s$ and between 25 and 38 $MT/yr$ of sediment.   
The delta is a product of the diversion of the Mississippi River in the 
1970s and since then it has evolved with minimum human alteration. 
We utilize the outline of the Wax Lake delta structure processed 
by~\citet{Edmonds11}; it has 59 links and 24 shoreline outlets. 
The partition of the flow at a node among the immediate downstream channels 
is proportional to the channel width ~\citep{BP03,Edmonds11}. 
The Niger delta is an older, highly complex distributary network 
that contains numerous loops and other intricate structures.  
We consider the area outlined by~\citet{SmartMoruzzi72}; it has
180 links and 15 shoreline outlets.  
In absence of the channel width data, we use equal partition 
of the flow among the immediate downstream channels.

\begin{figure}
\noindent\includegraphics[width=.5\textwidth]{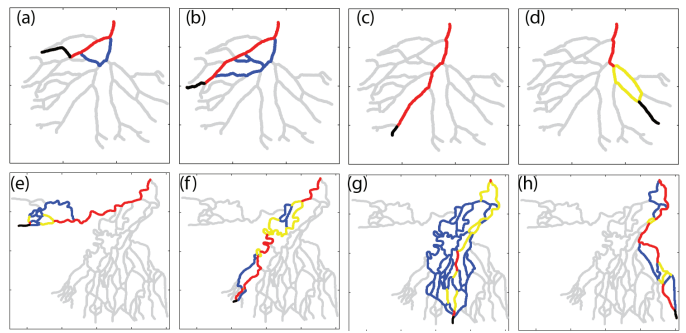}
\caption{Vulnerability maps for the Wax Lake (a-d) and Niger (e-h) deltas.  
Each panel highlights the contributing network for a single outlet. 
Shoreline outlets are shown in black.  
Red, yellow, and blue links represent high($r>$40\%), medium (20\%$< r <$40\%) and low ($r<20\%$) reduction to the shoreline outlet, respectively, where a 40\% flux reduction is applied to the link.}
\label{fig:fig3}
\end{figure}

The steady state flux for the Wax Lake delta is illustrated in Fig.~\ref{fig:fig2}(a). 
There exists no dominant shoreline outlet for this delta --
the maximum outlet flow of about 12\% of the apex flux is achieved at 4 out of 24
outlets.
The flux at the shoreline shows that 25\% (6 out of 24) of outlet links 
receive 60\% of the flux at the apex.  
This result is comparable to the synthetic sediment flux distribution at 
the shoreline for the Wax Lake delta obtained by~\citet{Edmonds11}.
The steady state flux for the Niger delta is illustrated in Fig.~\ref{fig:fig2}(c).
This delta has a singe dominant outlet that receives 50\% of the apex flux,
with the second largest outlet receiving 12\%. 
Figures~\ref{fig:fig2}(b),(d) show the number of outlets a given link contributes to in
the two examined deltas. 
This plot highlights the relative importance of a link in the delta network: 
the \emph{hotspot} (red) links affect many outlets while blue links 
only affect a single outlet.  
The \emph{hotspots} can be interpreted as ``highways of perturbation'' since, 
even if the steady flux in the link is not high, the effect of the perturbation 
will be experienced by many shoreline outlets in the delta.  
Representative examples of the outlet contributing networks are highlighted 
in Fig.~\ref{fig:fig3}(a)-(d) (where the colors should be ignored for now).
In the Wax Lake delta, most of the outlet contributing networks 
(18 out of 24, or 67\%) have a single path connecting the delta apex 
to the shoreline outlet.
On the contrary, in the Niger delta all the networks have multiple pathways. 

Flux reduction is recurrent in deltas due to dams and impoundments.  
Flux reduction at edge $(uv)$ leads to flux reduction within 
the nourishing area of $v$.
The flux reduction is specified by adding
a new outlet $z$ to the vertex $v$ and assigning new weights 
$w_{uz}=w_{uv}^{\rm old}(1-\alpha)$ and 
$w_{uv}^{\rm new}=w_{uv}^{\rm old}\alpha$, $0<\alpha<1$.
The reduced flux in the entire delta is computed using Proposition~\ref{SF}
with the updated weighted adjacency matrix.
We are interested in identifying ``vulnerable'' links defined as the links 
whose flux reduction would cause the highest reduction at the outlets.
For that we compute the flux reduction at the outlets 
caused by an $\alpha = 0.4$ flux reduction at a given link,
considered one-by-one.  
The results are illustrated in \emph{vulnerability maps} for the representative 
subnetworks in Fig.~\ref{fig:fig3}.    
The links colored red, yellow, and blue represent high ($r >40\%$), 
medium ($20\% < r < 40\%$) and low ($r<20\%$) flux reduction at the 
shoreline outlet, respectively. 
In general, if a given link drains $p \cdot 100\%$, $0\le p\le 1$, of its flux to a given outlet, 
and the steady flux $f^{\rm old}$ at the link is related 
to the steady flux $g^{\rm old}$ at the outlet as $f^{\rm old}=Cg^{\rm old}$, 
$C>0$, then $\alpha$-reduction at this link results
in the outlet reduction $g^{\rm new} = f^{\rm old}(1-\alpha\,p\,C)$.
Here, $\alpha$ is a local characteristic of the link, while
both $C$ and $p$ are spatially extended pairwise characteristics of
the link and the examined outlet.   
We also notice that in multi-paths networks (Fig.~\ref{fig:fig3} except (c)) 
the vulnerability might not be monotone along individual downstream channels.
These observations support the necessity of a systematic, spatially-extended 
approach to studying the effects of link modifications.

In contrast to well-studied topology of tributary channel networks 
(networks that drain to a single outlet; e.g. \citep{RodriguezIturbe97}), 
the exploration of the topology of distributary channel networks (networks that originate from 
a single source and drain to multiple outlets) is still in its infancy.  
Yet, this topology defines the distribution of network fluxes and dictates how changes 
in a given part of a network propagate to the rest; it also paves the way to better understand the intricate self-organization of deltaic systems.   
We present here a framework for analyzing the topology of delta networks, and specifically for identifying the upstream and downstream subnetworks for any given vertex, computing steady-state flux propagation in the network, and performing vulnerability analysis by assessing parts of the network where a change would most significantly affect the downstream or shoreline fluxes.
Notably, all the results follow directly from the spectral decomposition of the Laplacian for
the graph representing the examined delta.
Although we focus on the steady state topology of a delta that does not directly incorporate the dynamic evolution of the channel morphology or its topology, extension of the framework to incorporate time-evolving adjacency matrix is possible. 
The proposed framework can form a basis for delta network topology classification and for defining comparative vulnerability metrics among different deltas as well as for the same delta under natural and/or human-induced changes.


\begin{acknowledgments}

This work is part of the BF-DELTAS project on ``Catalyzing action towards sustainability of deltaic systems'' funded by the Belmont Forum and the forthcoming 2015 ``Sustainable Deltas Initiative'' endorsed by ICSU.  The research is also supported by the FESD Delta Dynamics Collaboratory EAR-1135427 and NSF grant EAR-1209402 under the Water Sustainability and Climate Program. 

\end{acknowledgments}

\bibliography{Delta_REVTEX}

\end{document}